\documentclass[prb,aps,twocolumn,showpacs,amsmath,amssymb]{revtex4}
\usepackage{graphicx,amsmath,bm}
\usepackage{dcolumn}

\begin{document}

\title{Hall Coefficient of Equilibrium Supercurrents Flowing inside Superconductors}

\author{Takafumi Kita}
\affiliation{Department of Physics, Hokkaido University, Sapporo 060-0810, Japan}
\date{\today}

\begin{abstract}
We study augmented quasiclassical equations of superconductivity 
with the Lorentz force, which is missing from the standard Ginzburg-Landau and Eilenberger equations.
It is shown that the magnetic Lorentz force on equilibrium supercurrents 
induces finite charge distribution and 
the resulting electric field to balance the Lorentz force.
An analytic expression is obtained for the corresponding Hall coefficient
of clean type-II superconductors with simultaneously incorporating the Fermi-surface and gap anisotropies.
It has the same sign and magnitude at zero temperature as the normal state
for an arbitrary pairing, having no temperature dependence specifically for the $s$-wave pairing.
The gap anisotropy may bring a considerable temperature dependence in the Hall coefficient and 
can lead to its sign change as a function of temperature,
as exemplified for a model $d$-wave pairing with a two-dimensional Fermi surface.
The sign change may be observed in some high-$T_{c}$ superconductors.

\end{abstract}

\pacs{03.75.Kk, 67.40.Db, 05.20.Dd}

\maketitle

\section{Introduction}

Einstein \cite{Einstein1905} pointed out in 1905 that the Lorentz force in electromagnetic fields
can be deduced naturally 
from the self-evident force on a charge at rest in an electric field 
with his theory of special relativity.
He has thereby provided a firm logical ground on the magnetic part of causing a deflection.
However, this force is absent in the modern theoretical accomplishments of superconductivity,
i.e.\ the standard Ginzburg-Landau equations \cite{GL,Werthamer69} 
and the quasiclassical Eilenberger equations \cite{Eilenberger68,Rainer83,LO86,Kopnin01}
derived microscopically from the Gor'kov equations.\cite{Werthamer69,Rainer83,LO86,Kopnin01}
Thus, our understanding on the magnetic Lorentz force in superconductors has remained
at a somewhat phenomenological level.
We here wish to make an improvement on this fundamental issue,
focusing our attention on equilibrium cases.

London \cite{London50} included the Lorentz force as a necessary ingredient
in his phenomenological equations of superconductivity.
They predict that an equilibrium supercurrent ${\bf j}_{s}\equiv
en_{s}{\bf v}_{s}$ 
in a magnetic field ${\bf B}$ accompanies an electric field:
\begin{equation}
{\bf E} = \frac{m}{2e}{\bm\nabla} v_{s}^{2}=\frac{1}{n_{s}ec}{\bf B}\times {\bf j}_{s},
\label{E-p1}
\end{equation}
with $m$ as the electron mass, $e$ $(<\!0)$ as the charge, ${\bf v}_{s}$ as the superfluid velocity,
$n_{s}$ as the superfluid density, and $c$ as the light velocity.
The second equality results from
the London equation ${\bm\nabla}\times {\bf v}_{s}=-(e/mc){\bf B}$ with the condition
$({\bf v}_{s}\cdot{\bm\nabla}) {\bf v}_{s}={\bf 0}$.
The expression implies that one could estimate the superfluid density $n_{s}$
through the Hall coefficient $({n_{s}ec})^{-1}$,
which would diverge towards the transition temperature $T_{c}$.
On the other hand, van Vijfeijken and Staas \cite{VS64} presented phenomenological two-fluid equations
with the Lorentz force, which modify Eq.\ (\ref{E-p1}) into
\begin{equation}
{\bf E} = \frac{n_{s}}{n}\frac{m}{2e}{\bm\nabla} v_{s}^{2}
=\frac{1}{nec}{\bf B}\times {\bf j}_{s},
\label{E-p2}
\end{equation}
where ${n}$ is the electron density.
Thus, the Hall coefficient 
is predicted to stay constant up to $T_{c}$
contrary to the London theory.
These considerations with the free-electron dispersion were extended by
Adkins and Waldram \cite{AW68} 
to incorporate the electronic band structure
from a somewhat different context of the Bernoulli potential,
with no explicit connection to the Lorentz force.
Specifically, they considered how a uniform supercurrent at $T=0$ modifies the 
Cooper pairing of non-spherical Fermi surfaces
to present an expression of the Hall coefficient,
which can take either sign just as the one of the normal state. 
Hong \cite{Hong75} and 
Omel'yanchuk and Beloborod'ko \cite{OB83} later performed
microscopic calculations of the equilibrium electric field due to an almost uniform supercurrent,
also with no direct relevance to the Lorentz force.
Using the Gor'kov equations with the free-electron density of states,
they obtained an expression in favor of Eq.\ (\ref{E-p2})
together with an additional term.
However, all the finite-temperature effects in
their derivations originate from the subtle energy dependence of the 
free-electron density of states,
so that they might be deduced to vanish for a constant density-of-states
near the Fermi level.
It should be noted finally that no investigations seem to have been carried out
for the cases of anisotropic pairings.

Pioneered by Onnes and Hof in 1914, 
efforts have also been made to detect
an equilibrium/quasi-equilibrium Hall voltage
of superconductors.\cite{Onnes14,Lewis53,JS59,Meservey65,BK68,MB71}
One can show with Eq.\ (1) or (2), 
the Maxwell equation ${\bm\nabla}\times {\bf B}=(4\pi/c){\bf j}_{s}$,
and the condition $({\bf B}\cdot{\bm\nabla}){\bf B}={\bf 0}$
that the Hall voltage $V_{H}$ in the Meissner state 
between the surface of the sample and its interior is given by
\begin{equation}
V_{H}=\frac{cR_{H}}{8\pi}B_{0}^{2},
\end{equation}
with $R_{H}$ the Hall coefficient and $B_{0}$ the external field.
It could be detected with a spheroid sample in a longitudinal magnetic field
by measuring the potential difference between a point on the equator and a pole;\cite{comment0}
see Refs.\ \onlinecite{Lewis53} and \onlinecite{BK68} for the experimental setup.
However, early experiments \cite{Onnes14,Lewis53,JS59,Meservey65} observed null Hall voltage
contrary to the theoretical predictions.
Hunt \cite{Hunt66} and
Nozi\`eres and Vinen \cite{NV66} later pointed out independently that voltmeters used in those experiments,
which require direct contacts to the sample, are not 
appropriate to detect the electrostatic potential.
Indeed, voltmeters can only pick out the chemical-potential difference, 
but the chemical potential is constant in equilibrium throughout the sample.
The difficulty was circumvented successively by applying capacitive couplings
to the specimen.\cite{BK68,MB71}
Bok and Klein \cite{BK68} performed a low-temperature measurement of the Hall voltage with Pb as well as
Nb and PbIn below $H_{c1}$
to obtain a good agreement of their results with Eq.\ (\ref{E-p1}).
Morris and Brown \cite{MB71} carried out a detailed experiment on Pb up to $T_{c}$
to report that their data point to Eq.\ (\ref{E-p2}) rather than Eq.\ (\ref{E-p1}).
However, detailed experiments over a wide range of materials still seem required to establish
the sign and the magnitude of the superconducting Hall coefficient 
in connection with the normal-state one.
Especially, no experiments seem to have been carried out on materials with
anisotropic energy gaps such as high-$T_{c}$ superconductors where new physics may be expected.

It was shown recently that 
the Lorentz force can be incorporated appropriately into the quasiclassical equations of 
superconductivity starting from the Gor'kov equations in the Keldysh formalism.\cite{Kita01}
The key procedures were: (i) an extension of the gauge-invariant Wigner transformation
introduced by Stratonovich \cite{Stratonovich56} and Fujita \cite{Fujita66} for the normal state 
to the Nambu Green's function; and (ii) a derivation of 
the corresponding Groenewold-Moyal product\cite{Groenewold46,Moyal49}
for performing the gradient expansion.
They have successfully removed the imperfect gauge invariance 
in a couple of preceding treatments.\cite{Kopnin94,HV98,Kopnin01}
The resulting equations can describe both the equilibrium and dynamical behaviors of 
superconductors with the Lorentz force such that
the normal-state Boltzmann equation is included appropriately as a limit.
Using them, we here develop a microscopic theory of the Lorentz force on equilibrium supercurrents
with the Fermi-surface and gap anisotropies.
We will thereby clarify: (i) the validity/applicability of 
the phenomenological results of Eqs.\ (\ref{E-p1}) and (\ref{E-p2}); 
and (ii) how the gap anisotropy affects them.
This step will also be necessary before elucidating dynamics of superconductors
microscopically where there still remain many unresolved issues
directly connected with the Lorentz force.\cite{NV66,BS65,Maki69,Ebisawa72,Dorsey92,KF95}

This paper is organized as follows. 
Section II presents the augmented quasiclassical equations of superconductivity
with the Lorentz force. 
Section III derives the expression of the Hall coefficient of equilibrium supercurrents.
Section IV presents its temperature dependence for both the $s$-wave
and $d$-wave pairings on a model two-dimensional Fermi surface.
Section V provides a brief summary.

\section{Augmented Eilenberger equations}

For simplicity, we first restrict ourselves to clean weak-coupling $s$-wave superconductors
in equilibrium.
The corresponding quasiclassical equations of superconductivity,
augmented so as to include the Lorentz force, are given by \cite{Kita01}
\begin{eqnarray}
&&\hspace{-7mm}
\left[\varepsilon \hat{\tau}_{3}-\hat{\Delta},\hat{g}^{{\rm R},{\rm K}}\right]+i\hbar{\bf v}_{\rm F}\cdot
{\bm\partial}\hat{g}^{{\rm R},{\rm K}}
\nonumber \\
&&\hspace{-7mm}
+\frac{i\hbar}{2}\!\left[ e{\bf v}_{\rm F}\cdot{\bf E}\frac{\partial}{\partial\varepsilon}
+\frac{e}{c}({\bf v}_{\rm F}\!\times\!{\bf B})\cdot
\frac{\partial}{\partial {\bf p}_{\rm F}}\right]\!\left\{\hat{\tau}_{3},
\hat{g}^{{\rm R},{\rm K}}\right\}=\hat{0}.
\label{QCE}
\end{eqnarray}
Here 
$\hat{g}^{{\rm R},{\rm K}}\!=\!\hat{g}^{{\rm R},{\rm K}}(\varepsilon,{\bf p}_{\rm F},{\bf r})$
are the $2\!\times\! 2$ retarded and Keldysh Green's functions, respectively, 
$\varepsilon$ denotes the excitation energy, 
$\hat{\tau}_{3}$ the third Pauli matrix, $\hat{\Delta}=\hat{\Delta}({\bf r})$
the gap matrix, ${\bf v}_{\rm F}$ the Fermi velocity, ${\bf p}_{\rm F}$ the Fermi momentum, 
$[\hat{P},\hat{Q}]\equiv\hat{P}\hat{Q}-\hat{Q}\hat{P}$,
and $\{\hat{P},\hat{Q}\}\equiv\hat{P}\hat{Q}+\hat{Q}\hat{P}$.
The quantity ${\bm\partial}$ denotes
${\bm\nabla}$, ${\bm\nabla}-i\frac{2e}{\hbar c}{\bf A}({\bf r})$,
or ${\bm\nabla}+i\frac{2e}{\hbar c}{\bf A}({\bf r})$ when operating on the diagonal, $(1,2)$, or $(2,1)$
element of $\hat{g}^{{\rm R},{\rm K}}$, respectively, 
with ${\bf A}({\bf r})$ the vector potential.
The advanced function $\hat{g}^{{\rm A}}$ is obtained from the retarded one by
$\hat{g}^{{\rm A}}=-(\hat{\tau}_{3}
\hat{g}^{{\rm R}}\hat{\tau}_{3})^{\dagger}$.

The term with ${\bf E}$ and ${\bf B}$ in Eq.\ (\ref{QCE})
represents the Lorentz force which is missing from
the Eilenberger equations.\cite{Eilenberger68,Rainer83,LO86,Kopnin01}
It is also absent in the standard Ginzburg-Landau equations \cite{GL,Werthamer69}
obtained from the Eilenberger equations as a limit.
Its relevance may be realized by taking the normal-state limit of 
$\hat{\Delta}\rightarrow \hat{0}$ and $\hat{g}^{{\rm R}}=-\hat{g}^{{\rm A}}=\hat{\tau}_{3}$;
then the $(1,1)$ element of Eq.\ (\ref{QCE}) for $\hat{g}^{{\rm K}}$ reduces to the
quasiclassical Boltzmann equation in static electromagnetic fields without the collision integral
and time dependence.
Thus, the term is indispensable for describing dynamical behaviors
of superconductors, and as seen below,
will also produce observable effects even in equilibrium.

The gap matrix in Eq.\ (\ref{QCE}) can be written as
\begin{equation}
\hat{\Delta}=\left[\begin{array}{cc}\vspace{1mm}
0 & -\Delta \\ \Delta^{*} & 0
\end{array}\right].
\label{hatDelta}
\end{equation}
Also considering the symmetry of Eqs.\ (72)-(75) in Ref.\ \onlinecite{Kita01}, 
we can express $\hat{g}^{{\rm R},{\rm K}}$ conveniently as
\begin{equation}
\hat{g}^{{\rm R}}=\left[\begin{array}{cc}\vspace{1mm}
g^{{\rm R}} & -if^{{\rm R}} \\ i\bar{f}^{\,{\rm R}} & -\bar{g}^{{\rm R}}
\end{array}\right],\hspace{5mm}
\hat{g}^{{\rm K}}=\left[\begin{array}{cc}\vspace{1mm}
g^{{\rm K}} & -if^{{\rm K}} \\ -i\bar{f}^{\,{\rm K}} & \bar{g}^{{\rm K}}
\end{array}\right],
\label{hatg^RK}
\end{equation}
where the barred functions are defined generally by 
\begin{equation}
\bar{g}^{{\rm R}}(\varepsilon,{\bf p}_{\rm F},{\bf r})\equiv 
[g^{{\rm R}}(-\varepsilon,-{\bf p}_{\rm F},{\bf r})]^{*}.
\label{gBar}
\end{equation}
The elements of $\hat{g}^{{\rm K}}$ 
further obey $g^{{\rm K}*}=g^{{\rm K}}$ and $\bar{f}^{\,{\rm K}*}=f^{{\rm K}}$.

Equation (\ref{QCE}) is supplemented by self-consistency equations for $\Delta$, 
${\bf B}$, and ${\bf E}$ to form a closed set of equations. 
They are given explicitly by \cite{Eilenberger68,Rainer83,LO86,Kopnin01}
\begin{equation}
\Delta \ln\frac{T}{T_{c}}
=\frac{1}{4i}\int_{-\infty}^{\infty} \left[\langle f^{\rm K}\rangle
-\frac{2i\Delta}{\varepsilon}\tanh\frac{\varepsilon}{2k_{\rm B}T}\right] d\varepsilon ,
\label{gapEq}
\end{equation}
\begin{equation}
{\bm\nabla}\times {\bf B}=\frac{4\pi}{c}{\bf j}_{s}, \hspace{6mm}
{\bf j}_{s}\equiv -\frac{eN(0)}{2}\int_{-\infty}^{\infty}  
\langle {\bf v}_{\rm F} g^{{\rm K}}\rangle d \varepsilon ,
\label{Maxwell1}
\end{equation}
\begin{equation}
{\bm\nabla}\cdot{\bf E}=4\pi\rho,\hspace{6mm}
\rho \equiv -\frac{eN(0)}{2}\int_{-\infty}^{\infty}  
\langle g^{{\rm K}}\rangle d \varepsilon,
\label{Maxwell2}
\end{equation}
where $\langle \cdots \rangle$ denotes the Fermi-surface average with $\langle 1 \rangle=1$,
$k_{\rm B}$ the Boltzmann constant,
and $N(0)$ the normal-state density of states per spin and unit volume at the Fermi level.
Equations (\ref{Maxwell1}) and (\ref{Maxwell2}) are just the Maxwell equations to determine the static
electromagnetic fields.

The gap anisotropy can be 
incorporated easily into the above formalism
by $\Delta({\bf r})\rightarrow
\Delta({\bf r})\phi({\bf p}_{\rm F})$ and $\langle f^{\rm K}\rangle
\rightarrow\langle f^{\rm K}\phi^{*}\rangle$ 
in Eqs.\ (\ref{hatDelta}) and (\ref{gapEq}), respectively, 
where $\phi$ is the basis function 
on the Fermi surface with $\langle|\phi|^{2}\rangle=1$.
Other possible extensions will be mentioned below near the end of 
Sec.\ III with respect to the Hall coefficient.

\section{Electric Field due to magnetic Lorentz force}

We embark on solving Eq.\ (\ref{QCE}) for the $s$-wave pairing 
by estimating the order of magnitude of the Lorentz force.
To this end, let us introduce the units where the energy is measured by the 
energy gap $\Delta_0$ at $T=0$ in zero fields, the length by 
$\xi_0\equiv \hbar \langle v_{\rm F}\rangle/\Delta_{0}$, 
the magnetic field by $B_{0}\equiv \hbar c/2|e|\xi_{0}^{2}$,
and the electric field by $E_{0}\equiv \Delta_{0}/|e|\xi_{0}$.
Dividing  Eq.\ (\ref{QCE}) by $\Delta_{0}$,
one may realize immediately that 
the magnetic Lorentz force in Eq.\ (\ref{QCE}) is 
an order of magnitude smaller in terms of $\delta\equiv \hbar/\langle p_{\rm F}\rangle \xi_{0}\ll 1$.
Since ${\bf E}$ is induced solely by the magnetic Lorentz force, as seen below, 
the term with ${\bf E}$ is also of the order of $\delta$.
It hence follows that we can carry out a
perturbation expansion of Eq.\ (\ref{QCE}) with respect to the
Lorentz force by expanding
\begin{equation}
\hat{g}^{{\rm R},{\rm K}}=\hat{g}^{{\rm R},{\rm K}}_{0}+\hat{g}^{{\rm R},{\rm K}}_{1}+\cdots .
\end{equation}
It is performed below up to the first order in $\delta$ to an excellent approximation.

We first neglect the Lorentz force in Eq.\ (\ref{QCE}) to obtain the equations of $O(\delta^{0})$.
They are just the standard Eilenberger equations
where the solutions $\hat{g}^{{\rm R},{\rm K}}_{0}$ satisfy
$\hat{g}^{{\rm R}}_{0}\hat{g}^{{\rm R}}_{0}=\hat{1}$, $\bar{g}^{\rm R}_{0}=g^{\rm R}_{0}$, and 
$\hat{g}^{{\rm K}}_{0}=(\hat{g}^{{\rm R}}_{0}
-\hat{g}^{{\rm A}}_{0})\tanh(\varepsilon/2k_{\rm B}T)$.\cite{Eilenberger68,Rainer83,LO86,Kopnin01}
The $(1,2)$ element of the equation for $\hat{g}^{{\rm R}}$ reads
\begin{equation}
-i\varepsilon f^{{\rm R}}_{0}+\frac{1}{2}\hbar {\bf v}_{\rm F}\cdot{\bm\partial} f^{{\rm R}}_{0}
=\Delta g^{{\rm R}}_{0} ,
\label{Eilenberger}
\end{equation}
with $g^{{\rm R}}_{0}=(1-f^{{\rm R}}_{0}\bar{f}^{\,{\rm R}}_{0})^{1/2}$,
which determines the whole solution.
Equation (\ref{Eilenberger}) with Eqs.\ (\ref{gapEq}) and (\ref{Maxwell1})
has been solved extensively to clarify vortex structures of 
$s$- and $d$-wave superconductors in equilibrium.\cite{KP73,Klein87,SM95,IHM99}

We next consider terms of $O(\delta)$ in Eq.\ (\ref{QCE}).
The corresponding $(1,1)$ and $(1,2)$ elements of the equation for 
$\hat{g}^{{\rm K}}_{1}$ read
\begin{subequations}
\begin{equation}
{\bf v}_{\rm F}\cdot{\bm\nabla}g^{{\rm K}}_{1}
-\frac{\Delta^{\! *} f^{{\rm K}}_{1}\!+\!\Delta \bar{f}^{\,{\rm K}}_{1}}{\hbar}
=-e {\bf v}_{\rm F}\cdot{\bf E}\frac{\partial g^{{\rm K}}_{0}}{\partial\varepsilon}
-\frac{e}{c} ({\bf v}_{\rm F}\times{\bf B})\cdot\frac{\partial g^{{\rm K}}_{0}}{\partial{\bf p}_{\rm F}},
\label{g^KEq1st}
\end{equation}
\begin{equation}
-i\varepsilon f^{{\rm K}}_{1}+\frac{1}{2}\hbar {\bf v}_{\rm F}\cdot{\bm\partial} f^{{\rm K}}_{1}
=\Delta \frac{g^{{\rm K}}_{1}-\bar{g}^{{\rm K}}_{1}}{2} .
\end{equation}
\label{f^KEq1st}
\end{subequations}
\hspace{-2.4mm}
The $(2,2)$ and $(2,1)$ elements are obtained from above by
setting $(\varepsilon,{\bf p}_{\rm F})\rightarrow (-\varepsilon,-{\bf p}_{\rm F})$, 
taking the complex conjugate,
and keeping Eq.\ (\ref{gBar}) and $\bar{g}^{{\rm K}}_{0}=-g^{{\rm K}}_{0}$ in mind.
The four equations determine $g^{{\rm K}}_{1}$,
$\bar{g}^{{\rm K}}_{1}$, $f^{{\rm K}}_{1}$, and $\bar{f}^{{\rm K}}_{1}$.
Writing them in terms of 
$g^{{\rm K}}_{1}+\bar{g}^{{\rm K}}_{1}$ and $g^{{\rm K}}_{1}-\bar{g}^{{\rm K}}_{1}$, 
we are led to linear closed equations for $g^{{\rm K}}_{1}-\bar{g}^{{\rm K}}_{1}$,
$f^{{\rm K}}_{1}$, and $\bar{f}^{{\rm K}}_{1}$ without the external source.
We hence conclude $f^{{\rm K}}_{1}=0$
and $\bar{g}^{{\rm K}}_{1}=g^{{\rm K}}_{1}$.
Substitution of this result into the equation for $g^{{\rm K}}_{1}+\bar{g}^{{\rm K}}_{1}$
yields
\begin{eqnarray*}
{\bf v}_{\rm F}\cdot{\bm\nabla}g^{{\rm K}}_{1}
=-e {\bf v}_{\rm F}\cdot{\bf E}\frac{\partial g^{{\rm K}}_{0}}{\partial\varepsilon}
-\frac{e}{c} ({\bf v}_{\rm F}\times{\bf B})\cdot\frac{\partial g^{{\rm K}}_{0}}{\partial{\bf p}_{\rm F}},
\end{eqnarray*}
which is clearly satisfied by the solution of
\begin{equation}
{\bm\nabla}g^{{\rm K}}_{1}
=-e {\bf E}\frac{\partial g^{{\rm K}}_{0}}{\partial\varepsilon}
-\frac{e}{c} {\bf B}\times\frac{\partial g^{{\rm K}}_{0}}{\partial{\bf p}_{\rm F}}.
\label{g^KEq1st2}
\end{equation}
We will use this latter equation below.

The same consideration for the equation of $\hat{g}^{{\rm R}}$ leads to the conclusion that
$f^{{\rm R}}_{1}=0$, $\bar{g}^{{\rm R}}_{1}=-g^{{\rm R}}_{1}$, and $g^{{\rm R}}_{1}$ 
is to be determined by Eq.\ (\ref{g^KEq1st2}) with the replacement 
$g^{{\rm K}}_{0,1}\rightarrow g^{{\rm R}}_{0,1}$.
However, the solution will not be necessary below
in the present clean limit.

To obtain a closed equation for ${\bf E}$, 
let us operate ${\bm\nabla}$ on Eq.\ (\ref{Maxwell2}).
We then approximate
$g^{{\rm K}}\approx g^{{\rm K}}_{0}+g^{{\rm K}}_{1}$,
substitute Eq.\ (\ref{g^KEq1st2}), and
use $\int_{-\infty}^{\infty}\langle g^{{\rm K}}_{0}\rangle d\varepsilon =0$
resulting from $\bar{g}^{{\rm K}*}_{0}=-g^{{\rm K}}_{0}$ as well as
$g^{{\rm K}}_{0}\rightarrow \pm 2$ for $\varepsilon\rightarrow \pm\infty$.
We thereby obtain
\begin{eqnarray}
-\lambda_{\rm TF}^{2}\nabla^{2} {\bf E}+{\bf E} = -\frac{{\bf B}}{4c}\times 
\int_{-\infty}^{\infty}\left<\frac{\partial g^{{\rm K}}_{0}}{\partial {\bf p}_{\rm F}}\right>
d\varepsilon ,
\label{Eq-E}
\end{eqnarray}
where $\lambda_{\rm TF}\!\equiv\! [8\pi e^{2}N(0)]^{-1/2}$ is the Thomas-Fermi screening length.\cite{FW71}

Equation (\ref{Eq-E}) is one of the main results of the present paper.
It enables us to calculate the induced electric field
of clean superconductors in equilibrium
with respect to the solution $g^{{\rm K}}_{0}$ of the standard Eilenberger
equations, i.e.\ Eqs.\ (\ref{Eilenberger}), (\ref{gapEq}), and (\ref{Maxwell1}).
Although derived above for the $s$-wave case,
Eq.\ (\ref{Eq-E}) is also valid in the presence of gap anisotropy, as seen below.
It implies that the electronic screening 
is the same in a superconductor as its normal state.
Since the source term on the right-hand side 
varies over the coherence length or the magnetic penetration depth 
which is much larger than $\lambda_{\rm TF}$,
we may generally neglect the first term on the left-hand side of Eq.\ (\ref{Eq-E})
to an excellent approximation.

Equation (\ref{Eq-E}) 
can be simplified further for the spherical Fermi surface with the slow-variation approximation.
Let us solve Eq.\ (\ref{Eilenberger}) perturbatively
up to the first order in terms of the gradient operator.
Putting the result
into $g^{{\rm R}}_{0}=(1-f^{{\rm R}}_{0}\bar{f}^{\,{\rm R}}_{0})^{1/2}$, 
we obtain
\begin{equation}
g^{{\rm R}}_{0}=\frac{-i\varepsilon}{W}
+\frac{\Delta^{\! *}{\bf v}_{\rm F}\cdot{\bm\partial}\Delta-
\Delta{\bf v}_{\rm F}\cdot{\bm\partial}\Delta^{\! *}}{4W^{3}},
\label{g^R-exp}
\end{equation}
where $W\equiv\sqrt{(-i\varepsilon)^{2}+|\Delta|^{2}}$, 
and an infinitesimal positive imaginary part is implied in $\varepsilon$.
We next substitute Eq.\ (\ref{g^R-exp}) into 
$g^{{\rm K}}_{0}\!=\!(g^{{\rm R}}_{0}\!-\! g^{{\rm A}}_{0})\tanh(\varepsilon/2k_{\rm B}T)$
and use it in Eqs.\ (\ref{Maxwell1}) and  (\ref{Eq-E}).
We then find
$$\int_{-\infty}^{\infty}
 \left<\frac{\partial g^{{\rm K}}_{0}}{\partial {\bf p}_{\rm F}}\right>d\varepsilon
=-\frac{6{\bf j}_{s}}{mN(0)v_{\rm F}^{2}e}
=-\frac{3c{\bm\nabla}\!\times\!{\bf B}}{2\pi mN(0)v_{\rm F}^{2}e}.$$
We further put this expression into Eq.\ (\ref{Eq-E}) together with $mN(0)v_{\rm F}^{2}=(3/2)n$
for the free-electron model.
Also neglecting the first term on the left-hand side, we obtain
$${\bf E}=\frac{1}{nec}{\bf B}\times{\bf j}_{s}.$$
Thus, Eq.\ (\ref{E-p2}) by van Vijfeijken and Staas is reproduced, i.e., the superconducting 
Hall coefficient is predicted to stay constant
up to $T_{c}$ 
for the $s$-wave pairing on the spherical Fermi surface, 
having the same sign and magnitude as that of the normal state.

Besides the Fermi-surface anisotropy, the gap aniso\-tropy can be 
incorporated easily into the above consideration
by $\Delta({\bf r})\rightarrow
\Delta({\bf r})\phi({\bf p}_{\rm F})$ and $\langle f^{\rm K}\rangle
\rightarrow\langle f^{\rm K}\phi^{*}\rangle$ 
in Eqs.\ (\ref{hatDelta}) and (\ref{gapEq}), respectively, 
where $\phi$ is the basis function on the Fermi surface with $\langle|\phi|^{2}\rangle=1$.
It is then straightforward to show that Eq.\ (\ref{Eq-E}) still holds, and
Eq.\ (\ref{E-p2}) with the slow-variation approximation is modified into
\begin{equation}
{\bf E}={\bf B}\times \underline{R}_{H}\,{\bf j}_{s}.
\label{E-BJ2}
\end{equation}
The corresponding Hall coefficient
is now a tensor:
\begin{equation}
\underline{R}_{H}=\frac{1}{2N(0)ec}
\left<\frac{\partial}{\partial {\bf p}_{\rm F}}(1-Y){\bf v}_{\rm F}\right>
\left< (1-Y){\bf v}_{\rm F}{\bf v}_{\rm F}\right>^{-1},
\label{R_H}
\end{equation}
where $Y\equiv Y({\bf p}_{\rm F},T)$ denotes the Yosida function\cite{Yosida58,Leggett75} given
in terms of $\varepsilon_{n}\equiv (2n+1)\pi k_{\rm B}T$ by
\begin{equation}
Y({\bf p}_{\rm F},T) \equiv 1- 2\pi k_{\rm B}T \sum_{n=0}^{\infty} 
\frac{|\Delta|^{2}|\phi({\bf p}_{\rm F})|^{2}}
{\bigl[\varepsilon_{n}^{2}+|\Delta|^{2} |\phi({\bf p}_{\rm F})|^{2}\bigr]^{3/2}}.
\label{Yosida}
\end{equation}
The factor $1-Y$ in Eq.\ (\ref{R_H}) acquires angular dependence for the anisotropic pairing
at finite temperatures due to the anisotropic distribution of thermally excited quasiparticles
embodied in $Y$.

We realize from Eq.\ (\ref{R_H}) with $Y({\bf p}_{\rm F},0) = 0$ that 
the superconducting Hall coefficient $\underline{R}_{H}$ 
at zero temperature should have the same sign and
magnitude for an arbitrary pairing as that of the normal state.
It agrees with the expression obtained by Adkins and Waldram at $T=0$.\cite{AW68} 
It is determined essentially by the integration of the curvature 
of the Fermi energy $\epsilon_{\rm F}\equiv
\epsilon({\bf p}_{\rm F})$ over the entire Fermi surface.
Especially, $\underline{R}_{H}$ has no temperature dependence
for the $s$-wave pairing where $1-Y$ in Eq.\ (\ref{R_H}) cancels.
In contrast, the gap anisotropy can bring a considerable
temperature dependence in $\underline{R}_{H}$,
as may be realized from $1-Y \propto |\phi|^{2}$ for $T\lesssim T_{c}$.
It is not ${\bf v}_{\rm F}$ itself for $T\lesssim T_{c}$ 
but ${\bf v}_{\rm F}|\phi|^{2}$ 
that is to be differentiated with respect to ${\bf p}_{\rm F}$.
In other words, the anisotropic distribution of
thermally excited quasiparticles also plays a crucial role for the
superconducting Hall coefficient at finite temperatures.
This will be demonstrated in Sec.\ IV by a model
calculation on a $d$-wave pairing.

We now consider several extensions.
When there are internal degrees of freedom in the relevant pairing,\cite{SU91} 
we need to change $\Delta({\bf r})\rightarrow
\sum_{i}\Delta_{i}({\bf r})\phi_{i}({\bf p}_{\rm F})$ 
in Eq.\ (\ref{hatDelta}) with $\langle \phi_{i}^{*}\phi_{j}\rangle = \delta_{ij}$
as well as $\Delta\rightarrow\Delta_{i}$ and
$\langle f^{\rm K}\rangle \rightarrow\langle f^{\rm K}\phi_{i}^{*}\rangle$ 
in Eq.\ (\ref{gapEq}).
It can be seen easily that Eqs.\ (\ref{Eq-E}) and (\ref{R_H}) still hold with 
a modification of
$|\Delta|^{2}|\phi|^{2}\rightarrow|\sum_{i}\Delta_{i}\phi_{i}|^{2}$ in Eq.\ (\ref{Yosida}).
The odd-parity case with ($\uparrow\downarrow,\downarrow\uparrow$) pairing
can be handled similarly with the modifications $\bar{f}\rightarrow-\bar{f}$
and $\phi^{*}_{i}\rightarrow -\bar{\phi}_{i}$
in the whole formulation.
This latter pairing was studied in terms of the superconductivity in Sr$_{2}$RuO$_{4}$
with a phenomenological Ginzburg-Landau functional to predict
a spontaneous Hall effect for a chiral $p$-wave state.\cite{FMS01}

We next consider the effects of impurities on the $s$-wave pairing
within the Born approximation for the $s$-wave scattering.\cite{Eilenberger68}
This is carried out by adding terms
$(i\hbar/2\tau)[\langle \hat{g}^{{\rm R}}\rangle, \hat{g}^{{\rm R}}]$ and
$(i\hbar/2\tau)\bigl(\langle \hat{g}^{\rm R}\rangle \hat{g}^{\rm K}
+\langle \hat{g}^{\rm K}\rangle \hat{g}^{\rm A}
-\hat{g}^{\rm R}\langle \hat{g}^{\rm K}\rangle
-\hat{g}^{\rm K}\langle \hat{g}^{\rm A}\rangle\bigr)$
on the left-hand side of Eq.\ (\ref{QCE}) for $g^{{\rm R}}$ and $g^{{\rm K}}$, respectively,
with $\tau$ denoting the relaxation time.
Then one can show that $g^{{\rm R}}_{1}$ still obeys
Eq.\ (\ref{g^KEq1st2}) with 
$g^{{\rm K}}_{0,1}\rightarrow g^{{\rm R}}_{0,1}$. 
On the other hand, the equation for $g^{{\rm K}}_{1}$ becomes more complicated
to prevent a straightforward extension of the clean-limit consideration.
Restricting ourselves to the Ginzburg-Landau region near $T_{c}$ and carrying out the 
expansion with respect to $\Delta$,\cite{Werthamer69} however,
one can show that:
(i) $g^{\rm K}_{1}\sim O(|\Delta|^{2})$ whereas $f^{\rm K}_{1}\sim O(|\Delta|^{3})$;
and (ii) $g^{\rm K}_{1}$ satisfies Eq.\ (\ref{g^KEq1st2}).
Thus, Eq.\ (\ref{E-p2}) is valid near $T_{c}$ even in the presence of impurities, and
also expected to hold approximately at lower temperatures.

We finally comment on the present results in terms of preceding theoretical treatments.
A transverse electric field is shown here to result naturally due to the magnetic Lorentz force,
in contrast to a treatment based on phenomenologically extended Ginzburg-Landau equations.\cite{KLB01}
Compared with those by Hong \cite{Hong75} and Omel'yanchuk and Beloborod'ko \cite{OB83}
for the free-electron model,
the present mechanism due to the Lorentz force requires 
no energy dependence in the density of states near the Fermi level,
thereby establishing the general existence of the transverse electric field among superconductors.
As for the additional contribution found by
Hong \cite{Hong75} and Omel'yanchuk and Beloborod'ko,\cite{OB83}
it is due to the energy dependence in the density of states, accompanied by
a reduction in the pair potential, 
and predicted to vanish at $T=0$.
Hence it may be distinguished clearly from Eq.\ (\ref{E-BJ2}) by experiments
on clean type-II superconductors in the Meissner state.
There is yet another mechanism of a finite electric field in superconductors
not directly connected with the supercurrent, i.e.,
that caused by a reduction in the pair potential such as the one in a
vortex core of type-II superconductors.\cite{KF95}
However, this effect can also be neglected for clean type-II superconductors in the Meissner state.

\section{Numerical Examples for the Hall coefficient}

To see the importance of the gap anisotropy on the equilibrium Hall coefficient,
we here present a model calculation of Eq.\ (\ref{R_H})
for a $d$-wave pairing.
We specifically consider the dimensionless
single-particle energy on a two-dimensional square lattice:
\begin{eqnarray}
&&\hspace{-10mm}
\epsilon_{\bf p}=-2(\cos p_{x}+\cos p_{y})+4t_{1}(\cos p_{x}\cos p_{y}-1)
\nonumber \\
&&\hspace{-2mm}
+2t_{2}(\cos 2p_{x}+\cos 2p_{y}-2) .
\label{E_k}
\end{eqnarray}
with $t_{1}=1/6$ and $t_{2}=-1/5$, which forms a band of 
$-4\leq \epsilon_{\bf p}\leq 4$. 
This model has been adopted
by Kontani and co-workers \cite{Kontani99,Kontani08} 
to describe the Fermi surface of cuprate superconductors
in theoretically investigating their normal-state Hall coefficients.
The Fermi surfaces for the average electron fillings $n=0.9$, $1.95$ per site are shown in Fig.\ 1.
Each of them is given in the extended zone scheme by a singly connected contour around 
$(p_{x},p_{y})=(\pi,\pi)$.

\begin{figure}[t]
\begin{center}
  \includegraphics[width=0.8\linewidth]{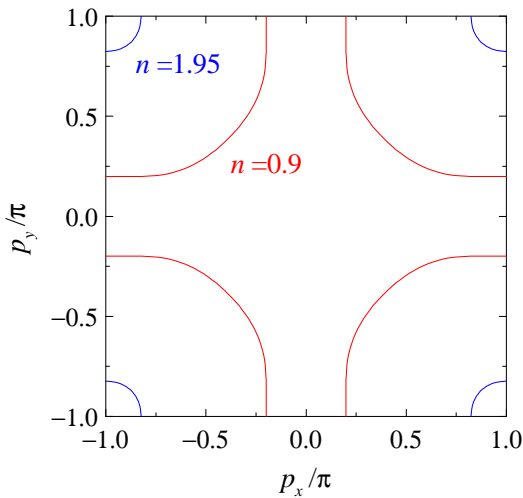}
\end{center}
  \caption{Fermi surfaces of $n=0.9$, $1.95$ for the 
  single-particle energy of Eq.\ (\ref{E_k}).}
  \label{fig:1}
\end{figure}

The normal-state Hall coefficient $R_{H}^{({\rm n})}\equiv R_{{\rm H}x}^{({\rm n})}=
R_{{\rm H}y}^{({\rm n})} $ for this band is obtained
by estimating Eq.\ (\ref{R_H}) with $Y=0$.
It is found that $R_{H}^{({\rm n})}$ changes its sign at the filling
$n_{c}=1.033$ ($\epsilon_{\rm F}=-0.121$) from negative to positive,
as shown in Fig.\ 2.
Thus, the Fermi surface for $n\sim n_{c}$ consists of competing portions with 
positive and negative curvatures which almost cancel with each other.
It is hence expected that the extra modulation of the curvature 
at finite temperatures due to the gap anisotropy,
embodied in the factor $1-Y$ of Eq.\ (\ref{R_H}), 
produces the most spectacular
effects around $n\sim n_{c}$.

It should be noted that the Fermi surface by Eq.\ (\ref{E_k}) is not sufficient 
to account for the signs and temperature dependences of the normal-state
Hall coefficient in high-$T_{c}$ superconductors, especially the positive sign of
$R_{H}^{({\rm n})}$ observed in Nd$_{2-x}$Ce$_{x}$CuO$_{4}$.\cite{Sato94}
Indeed, the vertex corrections due to the strong antiferromagnetic fluctuations
have been shown crucial for explaining the observed behaviors of $n\sim 1$.\cite{Kontani99,Kontani08}
However, we expect that the single-particle model adopted here will be sufficient
to capture the essential physics which the gap anisotropy brings
into the superconducting Hall coefficient.

\begin{figure}[t]
\begin{center}
  \includegraphics[width=0.85\linewidth]{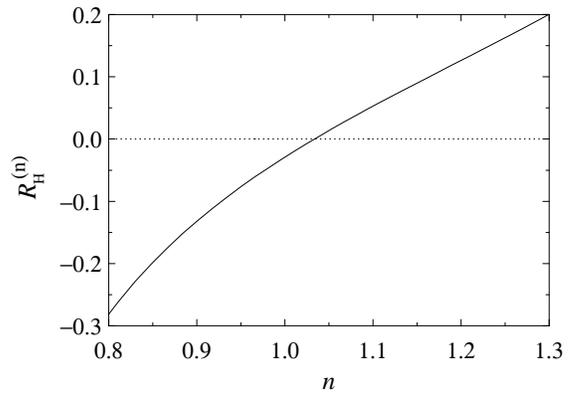}
\end{center}
  \caption{The normal-state Hall coefficient $R_{H}^{({\rm n})}$
  as a function of the filling $n$
  for the single-particle dispersion of Eq.\ (\ref{E_k}).}
  \label{fig:2}
\end{figure}

To see this, we here adopt a model $d$-wave pairing appropriate for $n\gtrsim 0.8$:
\begin{equation}
\phi({\bf p}_{\rm F})= A\bigl[(p_{{\rm F}x}-\pi)^{2}-(p_{{\rm F}y}-\pi)^{2}\bigr],
\end{equation}
where $A$ is the normalization constant determined by $\langle |\phi|^{2}\rangle =1$.

Figure 3 displays $R_{H}\equiv R_{{\rm H}x}=R_{{\rm H}y}$
of the equilibrium supercurrent
for $n=0.9$, $1.95$ calculated by Eq.\ (\ref{R_H}).
It is normalized by the normal-state Hall coefficient $R_{H}^{({\rm n})}$ for convenience.
With $n=1.95$ where the Fermi surface is almost isotropic and free-electron-like with positive charge,
$R_{H}$ increases monotonically from $R_{H}/R_{H}^{({\rm n})}=1.0$ at $T=0$ to 
$R_{H}/R_{H}^{({\rm n})}\sim 3.0$ at $T=T_{c}$.
On the other hand, $R_{H}/R_{H}^{({\rm n})}$ for $n=0.9$ 
even changes the sign as the temperature is increased from $T=0$.
These strong temperature dependences are brought about by the modulation of the Fermi-surface curvature
by the anisotropic distribution of thermally excited quasiparticles.

\begin{figure}[b]
\begin{center}
  \includegraphics[width=0.8\linewidth]{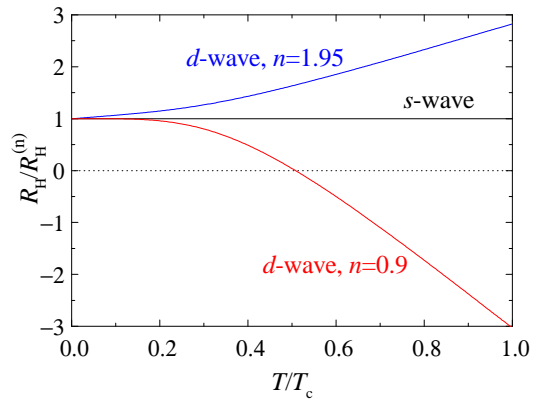}
\end{center}
  \caption{The Hall coefficient $R_{H}$ of equilibrium supercurrents,
  normalized by the normal-state coefficient $R_{H}^{({\rm n})}$,
  as a function of temperature for the $s$-wave pairing, the $d$-wave pairing with $n=0.9$, 
  and the $d$-wave pairing with $n=1.95$.}
  \label{fig:3}
\end{figure}

The sign change in the Hall coefficient as a function of temperature/magnetic field
has been observed in several high-$T_{c}$ superconductors in the vortex state with dissipative 
currents.\cite{Galffy88,Iye89,Artemenko89,Hagen90,Hagen91,Ong91,Luo92}
The origin of this sign change still remains mysterious,
because any attempt to analyze it theoretically necessarily has to clarify
complicated vortex motions of type-II superconductors
with electromagnetic fields.
On the other hand, we have shown here that the sign change can occur
even in the Hall coefficient of equilibrium supercurrents, which is much simpler
without vortex motions,
due to the modification of the Fermi-surface curvature at finite temperatures caused by
the anisotropic distribution of thermally excited quasiparticles.
It may be detected in some high-$T_{c}$ superconductors.
An observation of this sign change 
will provide: (i) an unambiguous support for the mechanism clarified here;
and (ii) a clue to understand the sign change in the vortex state
with dissipative currents.
We note in this context that neither the magnetic Lorentz force nor the gap anisotropy
were incorporated in the phenomenological theories on the vortex motion\cite{BS65,NV66,Hagen90,Hagen91,KF95}
and in the microscopic theory based on the time-dependent Ginzburg-Landau 
equations.\cite{Dorsey92}

\section{summary}

We have developed a microscopic theory of the Lorentz force in
equilibrium superconductors using a theoretical framework which
embraces the normal-state Boltzmann equation.
The magnetic Lorentz force working on equilibrium supercurrents is 
shown to induce an electric field as Eq.\ (\ref{E-BJ2}), 
which has the same expression as the normal-state one with dissipative currents.
Using the slow-variation approximation appropriate for type-II superconductors,
we have obtained an analytic expression for the Hall coefficient in the clean limit
as Eq.\  (\ref{R_H}).
It tells us that: (i) $R_{H}$ at $T=0$
carries the same sign and magnitude as that of the normal state;
(ii) the coefficient stays constant up to $T_{c}$ for the $s$-wave pairing;
and (iii) $R_{H}$ can have a considerable temperature dependence for a non-isotropic energy gap
due to the anisotropic quasiparticle distribution at finite temperatures.
We have shown in terms of the point (iii) that 
a sign change in $R_{H}$ may result, as seen in Fig.\ 3.
This sign change in the equilibrium  Hall coefficient may be observed 
in some high-$T_{c}$ superconductors in the Meissner state.
The present mechanism for the sign change, 
which has not been considered in any of the preceding treatments,
may also play an essential role in the sign change of $R_{H}$
observed in the vortex state with dissipative 
currents.\cite{Galffy88,Iye89,Artemenko89,Hagen90,Hagen91,Ong91,Luo92}

Further experiments for a wide range of materials 
are desired on the Hall voltage in the Meissner state
for probing the Lorentz force through the sign and magnitude of the superconducting Hall coefficient.
The electric field will also be present in the vortex-lattice state 
to form a long-range periodic pattern, which may in principle be detected by experiment.

The author would like to thank H.\ Kontani for useful discussions
on the Hall coefficient in the normal states of high-$T_{c}$ superconductors,
and M.\ Ido on the properties of high-$T_{c}$ superconductors.
This research is partly supported by Grant-in-Aid for Scientific Research 
from the Ministry of Education, Culture, Sports, Science, and Technology
of Japan.


\begin{thebibliography}{99}
\bibitem{comment0}In this geometry, the surface electrostatic potential is constant 
on the circle perpendicular to the magnetic field.
The magnitude decreases continuously from the equator towards the pole where
the potential is the same as the interior of the sample.
\end{thebibliography}
\end{document}